\documentclass[12pt]{iopart}
\usepackage{iopams}
\usepackage{graphicx}                
\usepackage{subfigure}       

\newcommand{\partiell}[2]{\frac{\partial^{#2}}{\partial {#1}^{#2}}}
\newcommand{\abl}[1]{\frac{d}{ d{#1} }}

\begin{document}

\title[Propagation of a probe pulse inside a BEC under conditions of
EIT]{Propagation of a probe pulse inside a Bose-Einstein condensate
  under conditions of electromagnetically induced transparency}

\author{Pablo Barberis-Blostein} 

\address{Instituto de
  Investigaciones en Matem\'aticas Aplicadas y en Sistemas,
  Universidad Nacional Autonoma de M\'exico,  Circuito Escolar s/n
  Ciudad Universitaria
  C.P. 04510
  M\'exico, D.F.
} 

\author{Omar Aguilar-Loreto} 

\address{Centro Universitario de la Costa Sur, Universidad de
  Guadalajara, Av. Independencia Nacional 151, Autl\'an, Jalisco,
  M\'exico}

\begin{abstract}
  We obtain a partial differential equation for a pulse travelling
  inside a Bose-Einstein condensate under conditions of electromagnetically
  induced transparency. The equation is valid for a weak probe pulse.
  We solve the equation for the case of a three-level BEC in $\Lambda$
  configuration with one of its ground state spatial profiles
  initially constant. The solution characterizes, in detail, the
  effect that the evolution of the condensate wave function has on
  pulse propagation, including the process of stopping and releasing
  it.

\end{abstract}
\maketitle
\section{Introduction}\label{sec:introduction}
\begin{figure}
  \subfigure[]{\label{fig:atomo}%
\includegraphics[width=0.5\linewidth]{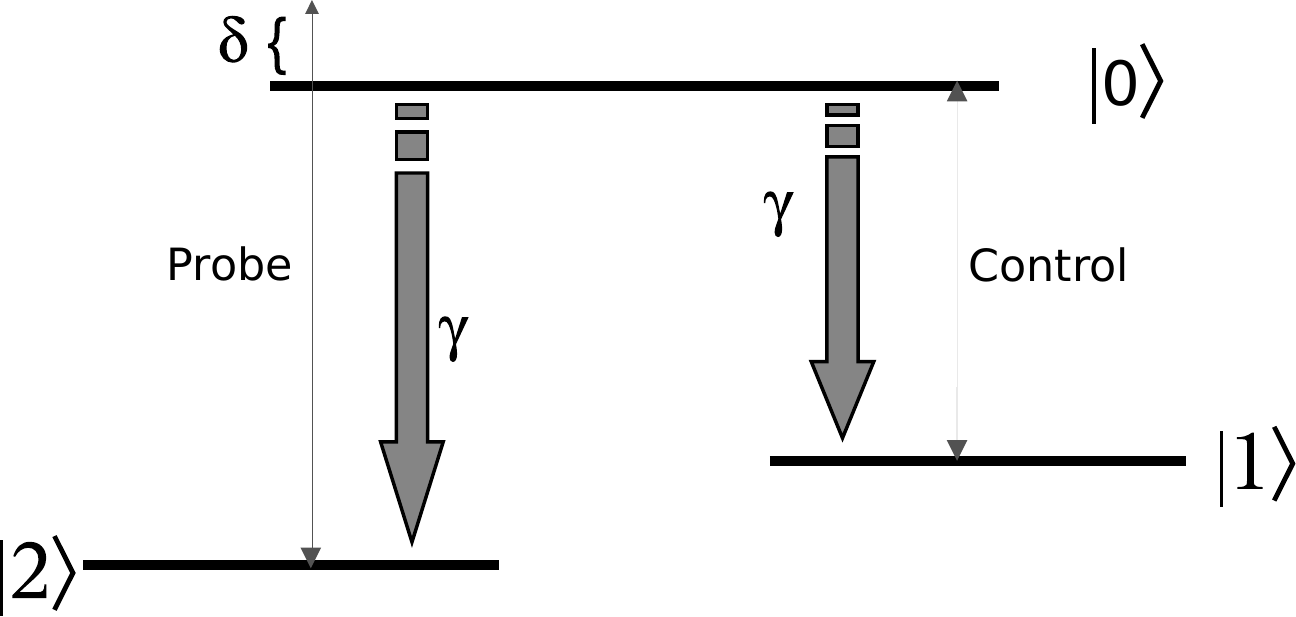}}
\quad
  \subfigure[]{\label{fig:absorption}%
\includegraphics[width=0.5\linewidth]{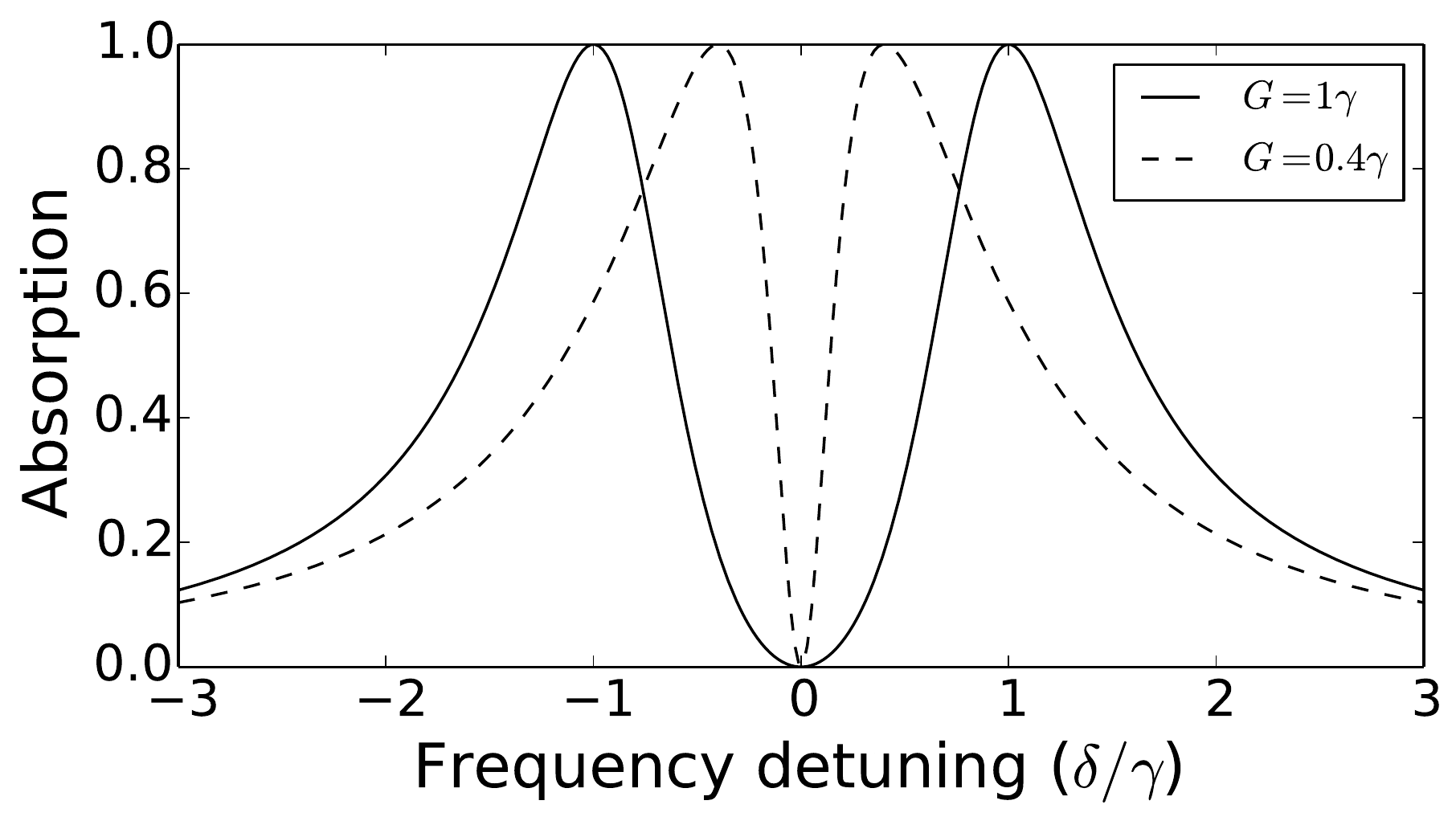}}
\caption{(a) Three-level atom in $\Lambda$ configuration interacting
  with two modes of the field. Transitions from state $|1\rangle$ to
  $|0\rangle$ and $|2\rangle$ to $|0\rangle$ are dipole
  allowed. Transitions from state $|1\rangle$ to $|2\rangle$ are not
  dipole allowed. The control mode of the field couples, resonantly,
  state $|1\rangle$ with $|0\rangle$. The probe mode of the field
  couples, with a detuning $\delta$, state $|2\rangle$ with
  $|0\rangle$. The rate of spontaneous decay from the excited state to
  the ground state is given by $\gamma$. (b) An example of the
  absorption profile for a weak probe field, with detuning $\delta$,
  interacting with three-level atoms. The interaction of the control
  field with the corresponding dipole transition is
  characterized by the Rabi frequency $G$. The absorption
  spectrum for two different values of $G$ are shown in the plot,
  note how the transparency window depends on the Rabi frequency. }
\end{figure}

A medium composed of three-level atoms in $\Lambda$ configuration (two
ground states, $1$ and $2$, dipole-coupled to an excited state $0$,
see Fig.~\ref{fig:atomo}) can be rendered transparent using a
technique known as electromagnetically induced transparency (EIT). We
will call the field mode coupling states $2$ and $0$ the probe field
and the field mode coupling states $1$ and $0$ the control field. In
the usual EIT configuration the probe field is weak compared with the
control field, which, in turn, is in resonance with the transition
frequency between states $1$ and $0$. We denote by $\delta$ the
detuning of the probe field from resonance with respect to the
transition frequency between states $2$ and $0$. The range of
frequency detunings, $\delta$, where absorption of the probe field by
the medium is negligible is known as the transparency window; its size
is proportional to the control field. The appearance of this
transparency window using the control field is known as
electromagnetically induced transparency; see
Fig.~\ref{fig:absorption} for an example of the typical absorption
spectrum of a probe field interacting with atoms showing EIT. In
Ref.~\cite{rv:marangos,rv:harris} the fundamentals of EIT are
explained in detail. The probe field inside a medium composed of
several three-level atoms exhibiting EIT propagates without distortion
with a velocity that is a function of the control field
\cite{rv:lukincollo,rv:rmpfleisch}; this can be used to slow down the
probe velocity as much as desired \cite{rv:hau,
  PhysRevLett.111.033601}.

For an atom, the atomic wave function has a dependence on its
position; when dealing with pulse propagation inside a medium composed
of atoms, we can ignore this dependence when the pulse size is much
larger than the variance of the atomic position; in this case only the
expectation value of the atomic position enters the description and we
say that we treat the atoms as point-like. The theory developed in
Ref.~\cite{rv:lukincollo,rv:rmpfleisch} treats the atoms as
point-like. When the atoms that compose the medium are in a
Bose-Einstein condensate, they can not longer be treated as
point-like. In this case the condensate wave function together with
its dynamics have a strong effect on the probe field propagation as is
shown through numerical simulations in
Ref.~\cite{Dutton2004,Dutton2002,Hau2008}. 
To observe this effect it is necessary that the
time during which the pulse interacts with the medium is larger than
the time scale characterizing the condensate wave function dynamics.

Since a
Bose-Einstein condensate has a well defined spatial phase relation,
the system evolution does not necessarily decohere the information
carried by the pulse, opening the possibility of using slow light and
BEC to manipulate optical information \cite{Hau2008,Riedl2013}.


The main question we address in this paper is: How does the evolution
of the condensate wave function modify the propagating pulse?
We address it in a form where analytical results can be obtained.
To solve this question we have to solve the Gross-Pitaevskii
equations, describing the condensate evolution, together with the
Maxwell equations, describing the field propagation. We obtain a
partial differential equation for the pulse propagation inside a
Bose-Einstein condensate with atoms in $\Lambda$ configuration. This
equation is much simpler than the set of equations that are usually
used to find the pulse propagation and the condensate wave function.
For some simple cases it can be solved analytically, showing
explicitly how the different elements -- probe pulse, control field
and wave functions -- interact and evolve. The equation is obtained
using a weak probe and approximations similar to those made in
Ref.~\cite{rv:memoria2}, but taking into account the condensate wave
function. For a condensate wave function with a spatial profile that
is initially constant, we write the solution in integral form. The
evolution for the field looks similar to the evolution of the wave
function for atoms in state $1$, but with its potential being described in a
reference system that moves with the pulse; we use this similarity for
analyzing the pulse propagation in the medium.

A general overview of a three-level Bose-Einstein condensate
interacting with two modes of the field is presented in
section~\ref{sec:model}, where known approximations are used to
express the model in a simplified way. In
section~\ref{sec:prop-equat-probe} we obtain the propagation equation
for the case of the probe pulse. In
section~\ref{sec:slow-light-inside} we study how the pulse propagates
inside the condensate showing EIT and discuss some physical
implications derived from the analytic solution. Concluding remarks
are given in section~\ref{sec:conclusion}.

\section{Model}\label{sec:model}
We only consider dynamics in the $x$ direction. The Hamiltonian of a three-level Bose-Einstein condensate interacting with two modes of light is
given by
\begin{equation}\label{eq:hamiltonian}
H=H_0+H_{int}\, ,
\end{equation}
with
\begin{eqnarray}
  H_0 &=& -\hbar \omega_{2} \int dx \, \hat{\psi}_2^{\dagger}(x)\hat{\psi}_2(x)-\hbar \omega_{1} \int dx \, \hat{\psi}_1^{\dagger}(x)\hat{\psi}_1(x)\nonumber\\
  & &+\sum_j \int dx \, \hat{\psi}_j^{\dagger}(x) \left ( -\frac{\hbar^2}{2 M}\frac{\partial^2}{\partial
    x^2} +V_j(x) \right ) \hat{\psi}_j(x)
  \nonumber\\ & & + \hbar \sum_j \int dx \, u_j(x) \hat{\psi}_j^{\dagger}(x)\hat{\psi}_j^{\dagger}(x)\hat{\psi}_j(x)\hat{\psi}_j(x) \nonumber\\
  & & + \frac{\hbar}{2} \sum_{i\neq j} \int dx \, u_{ij}(x) \hat{\psi}_j^{\dagger}(x)\hat{\psi}_i^{\dagger}(x)\hat{\psi}_j(x)\hat{\psi}_i(x)\, ,
  \nonumber\\
\end{eqnarray}
where $\hat{\psi}^{\dagger}_j(x),\hat{\psi}_j(x)$ are bosonic
operators that create and destroy a particle, with mass $M$, in the
state $j$, with $j=0,1,2$, at position $x$; these obey commutation rules
\begin{equation}\label{eqn:commutator}
\left [ \hat{\psi}_i(x),\hat{\psi}^{\dagger}_j(x') \right ] = \delta_{ij} \,
\delta (x-x')\, .
\end{equation}
The functions $V_{j}(x)$ are the trapping potential for the atoms in
state $j$. The constants $u_j$ are potentials due to elastic
collisions between particles in the same level $j$, $u_{ij}$ are
potentials due to elastic collisions between particles in level $i$
with particles in level $j$. We set the energy of the state $j=0$ as
the zero of the energy scale. The energy of atomic state $j=1,2$ is
given by $-\hbar \omega_j$ and the interaction Hamiltonian reads
\begin{eqnarray}
H_{int}&=&\hbar G(t) \int dx \,\left ( \hat{\psi}_0^{\dagger}(x)\hat{\psi}_1(x) e^{i (k_G x-\omega_L t)} +h.c. \right )
\nonumber\\& &+ \hbar g \int dx \,  \mathcal{E}(x,t)
\left ( \hat{\psi}_0^{\dagger}(x)\hat{\psi}_2(x) +h.c. \right )\, ,
\end{eqnarray}
where $G(t)$ is proportional to the strength of the continuous control
pulse, with wave number $k_G$ and frequency $\omega_L$,
$\mathcal{E}(x,t)$ is the incoming probe pulse, assumed to be centered
at a frequency $\bar{\omega}$, and $g$ is the atom-field coupling.

In order to calculate the system evolution we use the Heisenberg
representation, where the state vectors are time-independent and the
evolution is given by time-dependent operators that obey the
Heisenberg equations of motion. The Heisenberg
equations of motion for the bosonic operators are
\begin{equation}
\abl{t} \hat{\psi}_j(x,t)=\frac{1}{i \hbar } \left [ \hat{\psi}_j(x,t),H \right ]\, ,
\end{equation}
substituting the Hamiltonian $H$ given by Eq.~(\ref{eq:hamiltonian}) we
get the following differential equations for the bosonic operators
\begin{eqnarray}\label{eq:complete_system}
  \abl{t} \hat{\psi}_0 &=& \left ( \frac{i \hbar}{2M}\frac{\partial^2}{\partial
    x^2}
    -\frac{i}{\hbar} V_0 -2i u_0 \hat{\psi}^{\dagger}_0\hat{\psi}_0 - i u_{01}
    \hat{\psi}^{\dagger}_1\hat{\psi}_1 \right . \nonumber\\ & & \left .-i u_{02} \hat{\psi}^{\dagger}_2\hat{\psi}_2 
  \right )\hat{\psi}_0 -i G(t) e^{i (k_G x-\omega_L t)} \hat{\psi}_1 -i
  g \mathcal{E} \hat{\psi}_2 \, , \nonumber \\ 
  \abl{t} \hat{\psi}_1 &=& \left ( i\hbar\omega_1+\frac{i \hbar
      }{2M}\frac{\partial^2}{\partial
    x^2} -\frac{i}{\hbar} V_1 -2i u_1 \hat{\psi}^{\dagger}_1\hat{\psi}_1
    - i u_{12} \hat{\psi}^{\dagger}_2\hat{\psi}_2 \right .\nonumber \\ & & \left . 
    -i u_{10} \hat{\psi}^{\dagger}_0\hat{\psi}_0
  \right )\hat{\psi}_1  -i G(t) e^{i (k_G x-\omega_L t)} \hat{\psi}_0 \, ,\nonumber \\ 
  \abl{t}
  \hat{\psi}_2 &=& \left ( i\hbar\omega_2+ \frac{i \hbar}{2M}\frac{\partial^2}{\partial
    x^2} -\frac{i}{\hbar} V_2
    -2i u_2 \hat{\psi}^{\dagger}_2\hat{\psi}_2 - i u_{20} \hat{\psi}^{\dagger}_0\hat{\psi}_0
  \right . \nonumber\\ & & \left .-i u_{21} \hat{\psi}^{\dagger}_1\hat{\psi}_1 \right )\hat{\psi}_2 -i g \mathcal{E}^* \hat{\psi}_0\, .
\end{eqnarray}
Here we have omitted the time and space dependence of the operators and
variables and we assumed that $u_{ij}=u_{ji}$.  

We proceed now by writing the field operator as the sum of its
expectation value plus fluctuations~\cite{Dalfovo1999}
\begin{equation}
  \label{eq:sum_cnumber_operator}
\hat{\psi}_i=\psi_i(x,t)1\!\!1+\delta\hat{\psi}_i\, ,  
\end{equation}
where $1\!\!1$ is the identity operator and $\psi_i(x,t)=\langle
\hat{\psi}_i \rangle$ ( $\langle\cdots\rangle$ indicates the quantum
expectation value) is known as the mean field. The Gross-Pitaevskii
equations are partial differential equations where the fluctuations,
$\delta\hat{\psi}_i$, are not taken into account. Their solutions
give an approximation to the evolution of the mean field,
$\psi_i(x,t)$, whose squared norm gives the condensate spatial profile
of the atoms in level $i$. The Gross-Pitaevskii equations are valid
when the mean field dominates the fluctuations, this is the case when
the interactions between atoms are weak and the system temperature is
well below the transition temperature for Bose condensation.

Using Eq.~(\ref{eq:sum_cnumber_operator}) in
Eqs.~(\ref{eq:complete_system}) and neglecting fluctuations, we obtain
a set of coupled Gross-Pitaevskii Equations, they look similar to
Eqs.~(\ref{eq:complete_system}), but with functions $\psi_i(x,t)$
instead of operators $\hat{\psi}_i(x,t)$.

We assume that the control field is in resonance with the dipole
transition $1\rightarrow 0$ of atoms at rest, $\omega_1=\omega_L$. The
dominant frequency of the probe field is $\overline{\omega}$, assumed
to be in resonance with the dipole transition $2\rightarrow 0$ of
atoms at rest, $\overline{\omega}=\omega_2$. In order to eliminate
rapidly rotating terms we make the following transformations:
$\psi_1\rightarrow e^{i \omega_1 t}\psi_1$, $\psi_2\rightarrow e^{i
  \overline{\omega} t}\psi_2$, $\psi_0\rightarrow e^{-i \omega_L
  t}e^{-i \overline{\omega} t}\psi_0$ and $\mathcal{E}\rightarrow
\mathcal{E} e^{i \omega_2 t}$. All the atoms are initially in state
$2$, i.e. $\psi_0(x,t=0)=\psi_1(x,t=0)=0$. We use the weak probe approximation, $G\gg g\mathcal{E}$, and
assume that the wave functions $\psi_j$ can be expanded in powers of
$\Omega=g\mathcal{E}$:
\begin{equation}\label{eq:expansion_omega}
\psi_j=\psi_j^{(0)}+\psi_j^{(1)} \Omega+O(\Omega^2)\, .
\end{equation}

The expansion limits the validity of our results to the case where the
Rabi frequency of the probe field is much weaker than the the Rabi
frequency of the control field, this is the standard approximation in
EIT based slow light physics. 

To zeroth order we obtain
 \begin{equation}\label{eq:futuragp}
\abl{t} \psi_2^{(0)} =  \left ( \frac{i \hbar}{2M}\frac{\partial^2}{\partial
    x^2}
  -\frac{i}{\hbar} V_2 -2i u_2
  \psi^{*(0)}_2\psi^{(0)}_2 \right )\psi^{(0)}_2 \, ,
\end{equation}

and $\psi_1^{(0)}=\psi_0^{(0)}=0$.

At first order we get
\begin{eqnarray}
  \abl{t} \psi_0^{(1)} &=& \left(\frac{i \hbar }{2M}\frac{\partial^2}{\partial
    x^2}
    -\frac{i}{\hbar} V_0- i u_{02}
    \psi^{*(0)}_2\psi^{(0)}_2-i \Delta -\frac{\gamma}{2}\right)
  \psi_0^{(1)}  \nonumber \\  & &  -i G(t) e^{i k_G x} \psi_1^{(1)} -i g \mathcal{E}
  \psi_2^{(0)} \, ,\label{eq:psi0}\\ 
  \abl{t} \psi_1^{(1)} &=& \left(\frac{i
      \hbar}{2M}\frac{\partial^2}{\partial
    x^2} -\frac{i}{\hbar} V_1- i u_{12}
    \psi^{*(0)}_2\psi^{(0)}_2\right)\psi_1^{(1)} \nonumber\\ & &
  -i G(t) e^{-i k_G
    x} \psi_0^{(1)}\, ,\label{eq:expanionb} \\ 
  \abl{t} \psi_2^{(1)} &=& \left (
    \frac{i \hbar}{2M}\frac{\partial^2}{\partial
    x^2} -\frac{i}{\hbar} V_2 -2i u_2
    \psi^{*(0)}_2\psi^{(0)}_2 \right )\psi^{(1)}_2 \nonumber \\
  & & -i u_2
  \psi^{*(1)}_2\psi^{(0)}_2\psi^{(0)}_2\, ,
\label{eq:expanionc}                                     
\end{eqnarray}
where $\Delta=\omega_L-\omega_2$. We inserted spontaneous decay from
state $0$, at rate $\gamma$, phenomenologically in
Eq. (\ref{eq:psi0}). Since the probe pulse is weak, the population
of level $1$ of the condensate is small, this is the reason that
justify keeping only terms up to first order in $\Omega$ in
Eq.~(\ref{eq:expansion_omega}).

\section{Propagation Equation for the Probe Pulse}\label{sec:prop-equat-probe}
In the slowly varying approximation, the equation for the
probe electromagnetic field becomes \cite{lb:scully}
\begin{equation}
\left(\partiell{t}{} +c \partiell{x}{} \right )\mathcal{\overline{E}} =i g e^{-i k_F x}\langle\psi^{*}_2 (x,t) \psi_0(x,t)\rangle\, ,
\end{equation}
with $\mathcal{\overline{E}}= \mathcal{E} e^{i k_F x}$ being the
slowly varying part of the electromagnetic field with
$k_F=k(\bar{\omega})$. To first order in $\Omega$, the
dipole operator in the previous equation is
\[
\langle\psi^{*}_2 (x,t) \psi_0(x,t)\rangle\approx
\langle\psi^{*(0)}_2 (x,t) \psi^{(1)}_0(x,t)\rangle\, .
\]

Eq.~(\ref{eq:expanionb}) implies
\begin{equation}\label{eqn:psi2psi0}
\psi_0^{(1)}=\frac{i e^{i k_G x}}{G} \left ( \abl{t} -\frac{i
  \hbar}{2M}\frac{\partial^2}{\partial
    x^2} +\frac{i}{\hbar} V_1+i u_{12} \psi^{*(0)}_2\psi^{(0)}_2\right )\psi_1^{(1)}\, .
\end{equation}

We assume that $\psi_0^{(1)}$ reaches its stationary value very
quickly compared with the time scale evolution of the other
wave functions. Note that in the stationary regime the system reaches the
electromagnetically induced transparency regime and the population of
state $0$ is negligible. These two considerations allow us to write
Eq.~(\ref{eq:psi0}) as
\begin{equation}\label{eq:dip1}
\psi_1^{(1)}\approx -\frac{g\mathcal{\overline{E}} e^{i k_F x}}{G}
e^{-i k_G x}\psi^{(0)}_2\, ;
\end{equation}
this expression is equivalent to Eq.~(27) in Ref~\cite{rv:memoria2}.
We insert Eq.~(\ref{eq:dip1}) into Eq.~(\ref{eqn:psi2psi0}) and denote
the condensate wave function of level $2$, when there is no field, as
\[
\psi_2^{(0)}=\alpha(x,t)\, ,
\]
obtaining
\begin{eqnarray}\label{eqn:dipole2quant}
  \langle\psi^{*(0)}_2 (x,t) \psi^{(1)}_0(x,t)\rangle&=&-\frac{ig
    e^{i k_Fx}\alpha^*(x,t)}{G(t)} \partiell{t}{} \frac{\mathcal{\overline{E}}(x,t)
    \alpha(x,t)}{G(t)}\nonumber\\&& -\frac{g e^{i k_G x}\alpha^*(x,t)}{G^2(t)} \frac{\hbar
  }{2M}\frac{\partial^2}{\partial
    x^2}
  e^{i (k_F-k_G
    )x}\alpha(x,t)\mathcal{\overline{E}}(x,t)\\&& +\frac{g e^{i k_F
      x}\mathcal{\overline{E}}(x,t)|\alpha(x,t)|^2}{\hbar G^2(t)} V_1+\frac{g e^{i k_F
      x}\mathcal{\overline{E}}(x,t)u_{12}|\alpha(x,t)|^4}{G^2(t)}\nonumber\, .
\end{eqnarray}                                  
Substituting in the propagation equation we obtain
 

\begin{eqnarray}\label{eq:progen}
\left(\partiell{t}{} +c \partiell{x}{} \right )\mathcal{\overline{E}}(x,t) &=&-\frac{g^2
  \alpha^*(x,t)}{G(t)}\left\{\partiell{t}{}\frac{\mathcal{\overline{E}}(x,t)\alpha(x,t)}{G(t)} \right.\nonumber\\&&\left. -\frac{i \hbar}{2 M G(t)} \left[-(k_G-k_F)^2 \mathcal{\overline{E}}(x,t)
  \alpha(x,t)\right.\right.\nonumber\\ &&\left.\left. -i 2 (k_G-k_F)
  \partiell{x}{}\mathcal{\overline{E}}(x,t) \alpha(x,t)+\frac{\partial^2}{\partial
    x^2}\mathcal{\overline{E}}(x,t) \alpha(x,t) \right]\right.\nonumber\\&&\left.+\frac{i
  \mathcal{\overline{E}}(x,t)\alpha(x,t)}{\hbar G(t)} V_1+\frac{i
  \mathcal{\overline{E}}(x,t)u_{12}|\alpha(x,t)|^2\alpha(x,t)}{G(t)} \right\}\, ,\nonumber\\
\end{eqnarray}
which is one of our main results. The function
$\alpha(x,t)=\psi_2^{(0)}(x,t)$ is given by the solution of
Eq.~(\ref{eq:futuragp}) and does not depend on the field or the
condensate wave functions of the other atomic levels. 
In the following section we take advantage of this to obtain an
analytic expression for the solution in the particular case of an
initially spatially uniform condensate. This solution exemplifies how
the different terms on the right side of Eq.~(\ref{eq:progen})
contribute to the probe pulse evolution.

\section{Slow light inside a condensate}\label{sec:slow-light-inside}
In this section, we study how the pulse probe propagates inside the
condensate showing EIT. We will focus on the case where
$\alpha(x,t)=\alpha e^{i\mu t}$, with $\alpha$ a constant, is a
solution of the Gross-Pitaevsky equation for the state $2$ of the
condensate.  Spatially uniform condensates have been achieved
experimentally \cite{gaunt_bose-einstein_2013}.  The potential, $V_2$,
that generates this distribution, could be a square well; it can also
be a wide harmonic potential and the following discussion is valid in
the region near the center of the potential, where there is little
change in the density distribution \cite{pethick2002bose}. This case
avoids the effect that density change in atoms in level $2$ has
on pulse propagation; we are interested in the effect that state
evolution of atoms in level $1$ has on the pulse. This approximation
allow us to obtain an analytic expression for the solution that
completely characterizes both, the pulse propagation and the atomic state
evolution.
Defining $k_t=k_G-k_F$,
Eq.~(\ref{eq:progen}) reads
\begin{eqnarray}
\left( \frac{\partial }{\partial t}+c\frac{\partial }{\partial x}\right)
\mathcal{E}\left( x,t\right) &=&-\frac{g^2 \left\vert \alpha \right\vert ^{2}}{G\left( t\right) }\left(
\frac{\partial }{\partial t} \left( \frac{\mathcal{E}\left( x,t\right)
}{G\left( t\right) }\right) +i\mu \left( \frac{\mathcal{E}\left( x,t\right)
}{G\left( t\right) }\right)\right.\nonumber\\&&\left. +\frac{i \mathcal{E}(x,t)}{\hbar G(t)}
V_1(x)+\frac{i \mathcal{E}(x,t)u_{12}|\alpha|^2}{G(t)} \right
) \nonumber\\&&+\frac{i g^2 \left\vert \alpha \right\vert ^{2} \hbar}{ 2 M G^{2}\left( t\right) }\left[
  -k_{t}^{2}-i2k_{t}\frac{\partial }{\partial x}+ \frac{\partial
    ^{2}}{\partial x^{2}}\right] \mathcal{E}\left( x,t\right)\, .
\end{eqnarray}
To solve this equation we use a reference system that moves
with the probe pulse. With substitution
\begin{equation}
  \label{eq:cambio_variable}
  u=-\frac{x}{c}+\int_0^t\frac{G^2(\xi)}{g^2 \left\vert \alpha \right\vert ^{2}+G^2(\xi)}d\xi
\end{equation}
we obtain
\begin{equation}\label{eq:solconst}
\mathcal{E}(u,t)=\frac{G(t)e^{i\phi(t)}}{\sqrt{G^2(t)+g^2 \left\vert \alpha \right\vert ^{2}}}
\exp{\left (\int_0^t \frac{-i}{\hbar}\tilde{H}(\tau) d \tau\right )}\,\,\mathcal{E}(u,t=0)\, ,
\end{equation}
where
\begin{equation}
  \label{eq:fakehamiltonian}
  \tilde{H}=\frac{1}{(1+G^2(t)/g^2 \left\vert \alpha \right\vert ^{2})}\left ( \frac{\hbar^2}{2 M}(k_t+\frac{i}{c}\frac{\partial}{\partial
    u})^2 +V_1(u,t)\right )\, 
\end{equation}
and
\begin{equation}\label{eq:phidet}
  \phi(t)=(\mu+u_{12}|\alpha|^2)\left(\int_0^t\frac{G^2(\tau)}{G^2(\tau)+g^2 \left\vert \alpha \right\vert ^{2}} d\tau-t\right)\,\, .
\end{equation}

The solution for the field shown in (\ref{eq:solconst}), includes some
of the usual characteristic behavior of light traveling through media
showing EIT. The pulse velocity depends on the value of $G(t)$; as
$G(t)\rightarrow 0$ the pulse is stopped and, as can be seen from
expression (\ref{eq:dip1}), its information is transferred to the
atoms \cite{rv:lukincollo}. Note that if $k_t=0$ and the pulse is
already stopped ($G(t)=0$), the operator $\tilde{H}$, defined in
(\ref{eq:fakehamiltonian}), coincides with the Hamiltonian operator of
an atom trapped in the potential $V_1$. As shown in
relation (\ref{eq:dip1}), the wave function of atoms in state $1$ is given
by the probe field once the pulse is trapped; then, while the pulse is
trapped, the atomic wave function for atoms in state $1$ evolves in
terms of its own Hamiltonian. As solution (\ref{eq:solconst}) shows, the
profile of it gets written in the pulse once it is released
($G(t)\rightarrow \infty$). If $k_t\neq 0$, the photons transfer
momentum to the atoms that go from state $2$ to state $1$. This
precisely appears in (\ref{eq:fakehamiltonian}) as a displacement of the
momentum operator by $k_t$. This effect has been used to move a pulse
by physically moving the atoms in state $1$, and then releasing the
pulse once the atoms are in a new position \cite{rv:hau2condensados}.

It is clear from the solution that the pulse is modified as it travels
through the medium, and in the process of stopping it; the strength of
this modification depends on $G(t)$ and on the trapping potential for
atoms in state $1$. The behavior described above is the expected
behavior for trapping a pulse in a BEC \cite{Dutton2004}. The
advantage of the analytic expression for the field
(\ref{eq:solconst}), is that it tells us how the pulse evolve: as if
it were a quantum particle with a Hamiltonian given by
Eq.~(\ref{eq:fakehamiltonian}). Note that $\tilde{H}$ looks similar to
the Hamiltonian operator for atoms in state $1$ but multiplied by the
factor $1/(1+G^2(t)/g^2|\alpha|^2)$ and in a reference frame that
moves with the pulse; when the control field is zero, $u=x$ and
$\tilde{H}$ coincides with the Hamiltonian of atoms in state $1$. Due
to these facts we claim that: the way the probe pulse propagation is
affected by the evolution of the wave function of atoms in state $1$,
is given by $\tilde{H}$. With this interpretation, the coefficient
factor $1/(1+G^2(t)/g^2 \left\vert \alpha \right\vert ^{2})$ in
Hamiltonian (\ref{eq:fakehamiltonian}), gives the strength of how the
evolution of the atomic wave function affects the propagating pulse.

We will now explore some of the
physical insights that solution (\ref{eq:solconst}) exhibits.

The field, shown in (\ref{eq:solconst}), does not depend on $u_1$:
collisions between atoms in state $1$ does not affect the field
propagation; in the limit of weak probe field, the number of atoms
transferred to state $1$ is small and collisions between them are
negligible. 

The field has a global phase which depends on $\phi(t)$ (see
(\ref{eq:solconst}), (\ref{eq:phidet})), which in turn depends on term
$u_{12}$ that characterizes the strength of collisions between atoms
in state 2 with atoms in state 1. Pulses that propagate in the medium
with different control fields, $G(t)$, will have different phases; the
phase accumulation is maximum when the pulse is stopped, and it is
negligible when $G(t)$ is large. Depending on which parameters are
known, making these pulses interfere could be used to extract the
value of $u_{12}$, $\mu$ or $g$. If we make interfere a pulse that has
been stopped and then released, with a pulse that propagates at
constant velocity, we could extract information on the phase advance
due to the process of trapping and releasing the pulse.


For intermediate cases ($G(t)\neq 0,\infty$) the pulse is modified;
how it is modified depends on the wave function dynamics of atoms in
state $1$ and on the coupling between this dynamics and the pulse,
$1/(1+G^2(t)/g^2 \left\vert \alpha \right\vert ^{2})$. For example,
if $V_1$ is constant inside the region the pulse propagates, the width
of the wave function for state $1$ expands as a free particle, then
the probe pulse width expands as if it were a free particle with mass
$M\,(1+G^2(t)/g^2|\alpha|^2)$. In this case, not taking into account
that the field strength changes as a function of the control field,
the evolution of the field, in the process of trapping it, would be
similar to the evolution of a free particle wave function whose mass
change with time.

The pulse propagation can be manipulated
using $V_1$. For example, if $V_1=M\omega^2 u^2$, the center of the
pulse (in the moving frame) will behave as if it were a harmonic
oscillator with mass $M\,(1+G^2(t)/g^2|\alpha|^2)$ and frequency
$\omega/(1+G^2(t)/g^2|\alpha|^2)$. 


\section{Conclusion}\label{sec:conclusion}

When a probe pulse of light propagates through a three-level
Bose-Einstein condensate under conditions of electromagnetically induced
transparency, its shape is modified by the evolution of the condensate
wave function. The strength of this modification is a function of the
control field. This is in contrast with the case where atoms are
assumed point-like, where the probe pulse propagates without
modification.

We derived a partial differential equation whose solution gives the
probe pulse propagation inside a three-level Bose-Einstein condensate
under conditions of electromagnetically induced transparency. We obtained an
analytic expression when the initial profile for the atoms in state
$2$ is constant. The solution shows explicitly how different
parameters of the system -- trapping potential of the atoms, time
dependent control field, scattering lengths, atom-field coupling --
affect the pulse propagation and the evolution of the condensate wave
function. 
The form of the solution for the probe propagation,
expression (\ref{eq:solconst}), is similar to the formal solution of the
Schrodinger equation for the wave function of an atom in state $1$, but
with the Hamiltonian given in a reference frame that moves with the
pulse.


The solution allows us to see how to manipulate the pulse by choosing
different trapping potentials for atoms in state $1$ and different
control fields. For example, if the potentials are zero, the width of
the probe pulse expands in time as if it were a free quantum particle,
but with an expansion rate which is a function of the control
field. The propagated pulse acquires a global phase that depends on
the control field and the scattering length between atoms in state $1$
and $2$.  This phase could be measured by performing interference
experiments with pulses that propagate with different control fields.


We are confident that our results can help in the discussion of manipulating
light memories using condensates.
\section{Acknowledgments}
We thank Giovanna Morigi and Stefan Rist for useful discussions. Work
supported in part by DGAPA-UNAM grant PAPIIT IN103714.

\end{document}